\documentclass[12pt]{article}
\usepackage{a4wide}
\usepackage[dvips]{graphicx}
\usepackage{epsf}                        % to make figures
\usepackage{amssymb}
\usepackage{latexsym}                    % to make fancy symbols
\usepackage{amsfonts}                    % to make fancy symbols

\begin{document}

\rightline{FPAUO-10/02}
\rightline{CERN-PH-TH/2010-077}

\vspace{1.5truecm}

%%%%%%%%%%%%%%%%%
\centerline{\LARGE \bf Charged Particle-like Branes in ABJM}
%\vspace{.5cm}
%\centerline{\LARGE \bf in Brane-Antibrane Systems}
\vspace{1.3truecm}

\centerline{
    {\large \bf Norberto Guti\'errez${}^{a,b,}$}\footnote{E-mail address:
                                   {\tt norberto@string1.ciencias.uniovi.es}},
   {\large \bf Yolanda Lozano${}^{a,}$}\footnote{E-mail address:
                                  {\tt ylozano@uniovi.es}},
    {\large \bf Diego Rodr\'{\i}guez-G\'omez${}^{c,}$}\footnote{E-mail address:
                                  {\tt d.rodriguez-gomez@qmul.ac.uk}}                              
    }
                                                            
\vspace{.4cm}

\centerline{{\it ${}^a$Department of Physics,  University of Oviedo,}}
\centerline{{\it Avda.~Calvo Sotelo 18, 33007 Oviedo, Spain}}

\vspace{.4cm}
\centerline{{\it ${}^b$Theory Group, Physics Division, CERN,}}
\centerline{{\it CH-1211 Geneva 23, Switzerland}}

\vspace{.4cm}
\centerline{{\it ${}^c$Centre for Research in String Theory, Department of Physics,}}
\centerline{{\it Queen Mary, University of London,}}
\centerline{{\it Mile End Road, London E1 4NS, UK}}

\vspace{1truecm}

%%%%%%%%%%%%%%%%%
\centerline{\bf ABSTRACT}
\vspace{.5truecm}

\noindent
We study the effect of adding lower dimensional brane charges to the 't Hooft monopole, di-baryon and baryon vertex configurations in $AdS_4 \times \mathbb{P}^3$. We show that these configurations capture the background fluxes in a way that depends on the induced charges, and therefore, require additional fundamental strings in order to cancel the worldvolume tadpoles. The study of the dynamics reveals that the charges must lie inside some interval in order to find well defined configurations, a situation familiar from the baryon vertex in $AdS_5 \times S^5$ with charges. For the baryon vertex and the di-baryon the number of fundamental strings must also lie inside an allowed interval. Our configurations are sensitive to the flat $B$-field recently suggested in the literature. We make some comments on its possible role. We also discuss how these configurations are modified in the presence of a non-zero Romans mass.\\
\\
%{\it PACS:} 11.25.-w; 11.27.+d\\
%{\it Keywords:} Branes; Duality; Supergravity

\newpage

\section{Introduction}

In the last two years important progress has been made towards understanding three dimensional superconformal field theories and their relation to the $AdS_4/CFT_3$ correspondence. Elaborating on the pioneering work of \cite{BL,G}, Aharony, Bergman, Jafferis and Maldacena  (henceforth ABJM) proposed a theory conjectured to describe M2 branes probing a $\mathbb{C}^4/\mathbb{Z}_k$ singularity, where the orbifold acts with weights $(1,1,-1,-1)$ \cite{ABJM}. The near horizon geometry is then $AdS_4\times S^7/\mathbb{Z}_k$. The ABJM theory is an ${\cal N}=6$ quiver Chern-Simons-matter theory with gauge group $U(N)_k\times U(N)_{-k}$ and marginal superpotential. The high degree of SUSY requires the superpotential coupling to be related to the  CS level $k$. Since the latter renormalizes trivially, the ABJM theory is expected to be exactly conformal. Being the superpotential coupling proportional to $k^{-2}$, an appropriate large $k$ limit  $N^{1/5}<<k<<N$ allows for a weak coupling regime. In turn, in the gravity dual, the appropriate description is in terms of Type IIA string theory on $AdS_4\times\mathbb{P}^3$, with various RR fluxes \cite{ABJM}.

As standard in the $AdS/CFT$ correspondence, the chiral ring of the field theory is to be identified with the KK harmonics of the dual geometry. While these operators have scaling dimensions of order one, there are in addition chiral operators whose scaling dimension is of order $N$. Typically they correspond to non-perturbative states in the gravity side realized as branes wrapping calibrated cycles. In the context of the more familiar $AdS_5/CFT_4$ correspondence, a canonical example is the di-baryon \cite{GK,Berenstein:2002ke}, which corresponds to a wrapped 
D3-brane. This brane looks like a particle in $AdS_5$, and therefore, from its mass in global $AdS$, one can read off its anomalous dimension. Yet another example of such class of operators still in $AdS_5/CFT_4$ is the baryon vertex  \cite{Witten}. It corresponds to a D5-brane wrapping the whole internal space. Since it captures the $N$ units of 5-form flux, it requires $N$ fundamental strings ending on it. These strings can be thought of as external quarks (or Wilson lines in the fundamental representation) and this naturally suggests to identify this wrapped D5-brane with the $\epsilon$ tensor of the $SU(N)$ gauge group of the field theory.

The $AdS_4/CFT_3$ case is less understood in general. Concentrating on the ABJM case, various particle-like branes of the ${\cal N}=6$ $AdS_4 \times \mathbb{P}^3$ background were  already discussed in \cite{ABJM}. The $\mathbb{P}^3$ space has $H^{q}(\mathbb{P}^3)=\mathbb{R}$ for even $q\le 6$. Thus, it is possible to have $D0$, $D2$, $D4$ and $D6$ branes wrapping a topologically non-trivial cycle  \cite{ABJM, Gaiotto:2009mv}. While the $D2$ and $D6$ branes capture RR flux and develop worldvolume tadpoles, the $D0$ and $D4$ branes do not, and therefore should correspond to gauge-invariant operators. A subtle issue affecting these configurations was raised in \cite{AHHO}. Since the D4-branes wrap a non-spin manifold, they carry a half-integer worldvolume magnetic flux due to the Freed-Witten anomaly \cite{Freed:1999vc}. On the other hand, matching with the natural interpretation in field theory of such objects as di-baryon-like operators requires to switch a flat half-integer background $B$-field\footnote{The original argument supporting this $B$-field in \cite{AHHO} concerns a detailed analysis of the supergravity charges, while the analysis of the D4 worldvolume dynamics arises as a consistency check. For more details we refer to the original paper.}.  More generally, these wrapped branes act as sources to vector fields in $AdS_4$ arising from the reduction of RR potentials on topologically non-trivial cycles. In turn, vector fields in $AdS_4$ admit quantization with either of the two possible fall-offs at the boundary \cite{Witten:2003ya}, which amount to either a dynamical boundary gauge field or to a global current (discussions in this context have appeared recently in \cite{Imamura:2008ji, Imamura:2009ur, Klebanov:2010tj, Benishti:2010jn}). Since a definite quantization must be chosen, it follows that either magnetic or electric sources are forbidden for the corresponding bulk field \cite{Witten:2003ya}. This might shed some light on the role of the $B$-field. Indeed, coming back to the D4-branes, the quantization allowing for the D4-branes to exist should correspond to that where the $U(1)$'s are non-dynamical. Under that assumption, a determinant-like di-baryon dual operator would be gauge invariant by itself and it would have the right dimension to agree with the gravity result. On the other hand, the quantization dual to dynamical $U(1)$'s would forbid the D4-branes, which might suggest that no $B$-field is needed. However, a full understanding of this very important point is, at present, still lacking. 

Similar comments should hold for the remaining wrapped branes. It has been argued that the D0-brane corresponds to a di-monopole operator in the CFT side. The D6-brane, very much like the baryon vertex in $AdS_5$, requires $N$ fundamental strings ending on it. Its dual operator  should then naturally involve the $\epsilon$ tensor of the gauge theory. On the other hand, the D2-brane wrapped on the $\mathbb{P}^1\subset \mathbb{P}^3$ develops a tadpole that has to be cancelled with $k$ fundamental strings. The dual operator is a monopole 't Hooft operator, realized as a  ${\rm Sym}_k$ product of Wilson lines \cite{ABJM}. As mentioned above, as of today, there is no fully satisfactory understanding of the role of these branes and their dual operators.

The gravitational configurations described above admit a natural generalization by allowing  non-trivial worldvolume gauge fluxes. It is the aim of this paper to generalize the spectroscopy of wrapped branes by adding such non-trivial worldvolume gauge fields. To that matter, we will assume that suitable boundary conditions are chosen in each case such that the discussed branes are possible. These generalized configurations are of potential interest for some $AdS/CMT$ applications (see for instance \cite{FLRK,HLT}), for example as candidates for holographic anyons in ABJM, as discussed recently in \cite{Hartnoll:2006zb, KL}. 

Allowing for a non-trivial worldvolume gauge field has the effect of adding lower dimensional brane charges. This modifies how the branes capture the background fluxes in a way that depends on the induced charges, such that, in some cases, additional fundamental strings will be required to cancel the worldvolume tadpoles. From  that point of view, the generalized configurations are similar to holographic Wilson loops. We will see that the D2 and D6-branes do not differ much from the zero charge case, although they are stable only if the induced charges lie below some upper bound. 
This situation is familiar from the baryon vertex with magnetic flux in $AdS_5\times S^5$, studied in \cite{JLR}. 
In these cases the energy of the bound state increases with the charge that is being induced. However adding charges allows to construct  more general baryon vertex configurations. We will see that for the D6-brane the number of quarks that forms the bound state can be increased in this manner. From this point of view adding flux provides an alternative mechanism to that proposed in \cite{BISY} for modifying the number of quarks. In turn, the D4-brane with flux behaves quite differently from the fluxless case, since it will require fundamental strings ending on it as opposed to the vanishing worldvolume flux case. As we will see, the study of its dynamics reveals that the whole configuration is stable if the magnetic flux lies within a given interval, being maximally stable for an intermediate value, and reducing to free quarks at the boundaries.

The paper is organized as follows. In section 2 we generalize the particle-like brane configurations studied in \cite{ABJM} to include a non-vanishing magnetic flux proportional to the K\"ahler form of the $\mathbb{P}^3$. This magnetic flux induces lower dimensional charges in the brane worldvolume and, in some cases, modifies the charge of the tadpole.
In section 3 we study the dynamics of these configurations and show that there exist some bounds on the number of branes that can be dissolved in the worldvolumes. In section 4 we show how the previous configurations are modified in the presence of a non-zero Romans mass. Finally in section 5 we present our conclusions, where we try to relate the existence of the bounds for the magnetic flux to the stringy exclusion principle of \cite{MS}.

\section{Particle-like branes in $AdS_4\times \mathbb{P}^3$ with magnetic flux}

In this section we generalize the particle-like brane configurations in \cite{ABJM} to include a non-vanishing magnetic flux. We analyze the various brane charges that are dissolved as well as the charges of the different tadpoles induced. Following \cite{AHHO}, an important observation is that the dual gravity background might actually involve a non-vanishing but flat $B_2$ field. It is possible to argue for such a shift by noting that the D4-brane with minimal flux (it will turn out essential for the argument that this minimal flux has to be half-integer due to the Freed-Witten anomaly \cite{Freed:1999vc}) should be dual to a di-baryon. In order to review this argument, we will consider first the D4-brane case before turning to the D2 and D6 cases. 

Let us first start with a lightning review of the $AdS_4\times \mathbb{P}^3$ background while collecting some useful formulae.

\subsection{The background}

In our conventions the $AdS_4\times \mathbb{P}^3$ metric reads 
\begin{eqnarray}
ds^2&=&\frac{4\,\rho^2}{L^2}dx_{1,2}^2+L^2\frac{d\rho^2}{4\,\rho^2}+L^2ds_{\mathbb{P}^3}^2\nonumber\\
&=&L^2\Bigl( \frac14 ds^2_{AdS_4}+ds^2_{\mathbb{P}^3}\Bigr)\, ,
\end{eqnarray}
with $L$ the radius of curvature in string units,
\begin{equation}
L=\Bigl(\frac{32\pi^2 N}{k}\Bigr)^{1/4}
\end{equation}
This is a good description of the gravity dual to the $U(N)_k\times U(N)_{-k}$ CS-matter theory \cite{ABJM} when $N^{1/5}<< k << N$.

It is well-known that for $\mathbb{P}^3$ one has $H^{q}(\mathbb{P}^3)=\mathbb{R}$ for even $q$. Indeed, parameterizing the $\mathbb{P}^3$ as (e.g. \cite{Pope:1984bd})
\begin{eqnarray}
ds_{\mathbb{P}^3}^2=&&d\mu^2+\sin^2\mu\,\Big[ d\alpha^2+\frac{1}{4}\sin^2\alpha\, \big(\cos^2\alpha\,(d\psi-\cos\theta\, d\phi)^2+d\theta^2+\sin^2\theta\, d\phi^2\big)\nonumber \\
&&+\frac{1}{4}\cos^2\mu\, \big(d\chi+\sin^2\alpha\, (d\psi-\cos\theta\, d\phi)\big)^2\Big]
\end{eqnarray}
where
\begin{equation}
0\ge \mu,\,\alpha\ge \frac{\pi}{2}\, ,\quad 0\ge \theta \ge \pi\, ,\quad 0\ge \phi\ge 2\pi\, ,\quad 0\ge \psi,\, \chi\ge 4\pi\ ,
\end{equation}
there is a $\mathbb{P}^2$ at fixed $\theta$, $\phi$, and a $\mathbb{P}^1$ at $\mu=\alpha=\pi/2$ and fixed $\chi$, $\psi$. 

The K\"ahler form 
\begin{equation}
J=\frac{1}{2}d\mathcal{A}\, ,
\end{equation}
where $\mathcal{A}$ is the connection in
$ds^2_{S^7}=(d\tau + \mathcal{A})^2 + ds^2_{\mathbb{P}^3}$, which in our coordinates reads:
\begin{equation}
\mathcal{A}=\frac{1}{2}\sin^2\mu\, \Big(d\chi+\sin^2\alpha\,\big(d\psi-\cos\theta\, d\phi)\Big)\, ,
\end{equation}
satisfies 
\begin{equation}
\int_{\mathbb{P}^1}\, J=\pi\, ,\qquad \int_{\mathbb{P}^2}\, J\wedge J=\pi^2\, , \qquad \int_{\mathbb{P}^3}\, J\wedge J\wedge J=\pi^3\, .
\end{equation}
Therefore,
\begin{equation}
\frac{1}{6}\,J\wedge J\wedge J= d{\rm Vol}(\mathbb{P}^3) \qquad {\rm with} \qquad {\rm Vol}(\mathbb{P}^3)=\frac{\pi^3}{6}\, .
\end{equation}

The $AdS_4 \times \mathbb{P}^3$ background fluxes can then be written as
\begin{equation}
F_2=\frac{2L}{g_s}J\, ,\qquad F_4=\frac{6}{g_s\,L}\,d{\rm Vol}(AdS_4)\, ,\qquad F_6=\frac{6\, L^5}{g_s}d{\rm Vol}(\mathbb{P}^3)
\end{equation}
where
\begin{equation}
\label{dilatonL}
g_s=\frac{L}{k}\, , \qquad L^4 k = 32 \pi^2 N\, .
\end{equation}
The flux integrals read
\begin{equation}
\label{fluxes}
\int_{\mathbb{P}^3}\, F_6=32\, \pi^5\, N\, ,\qquad \int_{\mathbb{P}^1}\,F_2=2\pi\, k\, .
\end{equation}

\subsection{The di-baryon vertex}

Consider a D4-brane wrapping the $\mathbb{P}^2$ in $\mathbb{P}^3$. This brane lives at fixed $\theta$ and $\phi$, and since it does not capture any background fluxes it does not require any fundamental strings ending on it. 

Since the $D4$-brane wraps a $\mathbb{P}^2$, which is not a spin manifold, it should carry a half-integer worldvolume gauge field flux through the $\mathbb{P}^1\subset\mathbb{P}^2$, due to the Freed-Witten anomaly \cite{Freed:1999vc}. Given that the gauge-invariant quantity in the worldvolume is $\mathcal{F}=B_2+2\pi\alpha'\, F$, this half-integer worldvolume flux can be cancelled through a shift of $B_2$. This motivated \cite{AHHO} to include a flat $B_2$-field in the
dual IIA background:
\begin{equation}
\label{shift_B2}
B_2=-2\pi\, J
\end{equation}
which should be considered in addition to the fluxes discussed in the previous section.

We can now consider a more general configuration where we add extra worldvolume flux $F={\cal N}\,J$ on top of the $F_{FW}=J$ required to cancel the Freed-Witten anomaly, such that the total worldvolume flux is $F_T=(\mathcal{N}+1)\, J$ with even-integer quantization (that is, $\mathcal{N}\in 2\mathbb{Z}$ being $\mathcal{N}=0$ the minimal case). As noted above, the quantity appearing in the brane worldvolume action is the combination $\mathcal{F}=B_2+2\pi\, F_T$. Putting together the various definitions, we have $\mathcal{F}=2\pi\, F$, that is, the $B_2$ shift and the extra half unit of worldvolume flux cancel each other and we can effectively work as if we had no background $B_2$-field and $F=\mathcal{N}\, J$.

The DBI action is then given by:
\begin{eqnarray}
\label{DBI_D4}
S_{DBI}&=&-\frac{T_4}{g_{s}}\int d^5\xi \sqrt{-{\rm det}(g+2\pi F)}=
-\frac{T_4}{g_{s}}\int d^5 \xi \, \sqrt{|g_{tt}|} \sqrt{g_{\mathbb{P}^2}}\, \Bigl(L^4+2(2\pi)^2F_{\alpha\beta}F^{\alpha\beta}\Bigr)\nonumber\\
&=&-\frac{\pi^2\, T_4}{2\,g_s}\,\Bigl(L^4+(2\pi\mathcal{N})^2\Bigr)\,\int dt\, \frac{2\, \rho}{L}\, .
\end{eqnarray}
Therefore, for non-vanishing magnetic flux the mass of the D4-brane satisfies
\begin{equation}
\label{energyD4}
m_{D4} L=N+k\,{\cal N}^2/8
\end{equation}
From here we can see explicitly that in the minimal flux case, $\mathcal{N}=0$, the background $B_2$ cancels the half-integer worldvolume flux induced by the Freed-Witten anomaly, such that $m_{D4}\,L=N$; thus naturally admitting an interpretation as a di-baryon.

The D4-brane with magnetic flux captures the $F_2$ background flux through the coupling
\begin{eqnarray}
S_{CS}&=& \frac12 (2\pi)^2\, T_4 \int_{\mathbb{R}\times \mathbb{P}^2} P[F_2]\wedge F \wedge A=
2\,(2\pi)^2 T_4 \, k\,{\cal N}\int_{\mathbb{R}\times \mathbb{P}^2} J\wedge J\wedge A\nonumber\\
&=&k\,\frac{{\cal N}}{2}\, T_{F1} \int dt A_t
\end{eqnarray} 
Therefore $k\,{\cal N}/2$ fundamental strings are required to end on it in order to cancel the tadpole. Note that, due to the quantization condition for $\mathcal{N}$, this quantity is an integer number. Moreover, the magnetic flux also dissolves D2 charge through the coupling:
\begin{equation}
S_{CS}=2\pi\, T_4 \int_{\mathbb{R}\times \mathbb{P}^2} C_3\wedge F=\frac{{\cal N}}{2} \, T_2 \int C_3
\end{equation}
Thus, the number of fundamental strings is $k$ times the number of dissolved D2 branes. In fact, as we will see in the next subsection, a single D2-brane requires $k$ fundamental strings ending on it. Thus, from this perspective, the fundamental strings ending on the D4 are cancelling the tadpole due to the dissolved D2-branes.

We will see in the next section that the D4-brane with the $k\, {\cal N}/2$ attached F-strings is stable if the magnetic flux lies in an interval, reducing to $k\, {\cal N}/2$ radial fundamental strings stretching from the D4-brane to infinity, i.e. to free quarks, at both ends of the interval.

Given that $F$ is proportional to the K\"ahler form on the $\mathbb{P}^2$ it satisfies that $\int_{\mathbb{P}^2}F\wedge F={\cal N}^2\pi^2$. Therefore, it also induces D0-brane charge in the configuration, through the coupling:
\begin{equation}
S_{CS}=\frac12 (2\pi)^2\, T_4 \int_{\mathbb{R}\times \mathbb{P}^2}  
C_1 \wedge F\wedge F= \frac{{\cal N}^2}{8}\, T_0 \int_{\mathbb{R}} C_1\, .
\end{equation}
However, as noted in \cite{AHHO}, there are relevant higher curvature corrections \cite{Green:1996dd}
\begin{equation}
\label{Higher_Curv}
\Delta S\sim \int C\wedge e^{\mathcal{F}}\wedge \sqrt{\frac{\hat{\mathcal{A}}(T)}{\hat{\mathcal{A}}(N)}}\, ,
\end{equation}
where $\hat{\mathcal{A}}$ is the $A$-roof genus
\begin{equation}
\hat{\mathcal{A}}=1-\frac{\hat{p}_1}{24}+\frac{7\, \hat{p}_1^2-4\,\hat{p}_2}{5760}+\cdots
\end{equation}
and the Pontryagin classes are written in terms of the curvature of the corresponding bundle as
\begin{equation}
\hat{p}_1=-\frac{1}{8\pi^2}\, {\rm Tr}\, R^2\qquad \hat{p}_2=\frac{1}{256\, \pi^4}\,\Big(({\rm Tr}\, R^2)^2-2\, {\rm Tr}\, R^4\Big)
\end{equation}
The relevant term in (\ref{Higher_Curv}) is then
\begin{equation}
\Delta S=(2\pi)^4\, T_4\int C_1\wedge \frac{1}{48}\, (\hat{p}_1(N)-\hat{p}_1(T))=-\frac{1}{24}\, T_0\int C_1
\end{equation}
Thus, the total D0 charge is 
\begin{equation}
\label{D0_in_D4}
\Big(\frac{\mathcal{N}^2}{8}-\frac{1}{24}\Big)\, T_0\int C_1
\end{equation}

This equation shows that the D4-brane contains dissolved D0-brane charge even for the minimal flux allowed. Note that the term $k {\cal N}^2/8$ in (\ref{energyD4}) can be identified with ($L$ times) the mass of the extra ${\cal N}^2/8$ D0-branes dissolved in the worldvolume due to the non-vanishing magnetic flux. Therefore, (\ref{energyD4}) can be interpreted as the energy of a threshold BPS intersection of ${\cal N}^2/8$ D0-branes and a D4-brane. 
We should note however that if we want to study the dynamics of the D4-brane with fundamental strings attached in the probe brane approximation, we need to take the strings distributed uniformly on the D4. Therefore, the Killing spinors preserved by each one of the F1 strings will be different and all supersymmetries will be broken. Nevertheless, since both the wrapped cycle and the worldvolume flux are topologically non-trivial, we expect the system to be at least perturbatively stable.

By making all the F1 strings end in the same point, such that they preserve the same Killing spinor, we expect that a SUSY generalization in terms of a spike can be found. The problem of finding D4-brane spiky  solutions in $AdS_4\times \mathbb{P}^3$ has been addressed recently in \cite{KL}, although in the ansatz taken there the deformation of the D4-brane due to the electric field is not taken into account. It would be interesting to check if spiky solutions exist for both non-vanishing electric and magnetic fields.

\subsubsection{On boundary conditions and dual operators}

Given the topology of $\mathbb{P}^3$ it is possible to consider the KK reduction of the 5-form and 7-form respectively on $\mathbb{P}^2$ and $\mathbb{P}^3$ giving rise to vectors in $AdS_4$. As discussed in \cite{Witten:2003ya} and further elaborated in a similar context in \cite{Imamura:2008ji, Imamura:2009ur, Klebanov:2010tj, Benishti:2010jn}, the two fall-offs are possible in $AdS_4$. \footnote{It must be noted that, from an 11d perspective, the $U(1)$ fields discussed here are not related to a topological symmetry as in \cite{Klebanov:2010tj, Benishti:2010jn}, which makes them more subtle.} Choosing one or the other amounts to the dual $U(1)$ symmetry being gauged or not. In turn, from the bulk perspective, this is seen as electric-magnetic duality (the so-called $\mathcal{T}$-operation). It is possible to define a $\mathcal{S}$-operation such that their combined action forms an $SL(2,\mathbb{Z})$ algebra, which then connects \emph{different} boundary CFT´s. The action of such algebra is far from being understood. However, one particular implication would be that depending on the boundary conditions that are chosen the allowed sources are either the magnetic or the electric ones. From this point of view, one might argue that the quantization dual to dynamical boundary gauge fields forbids D4, D6 (which are electrically charged under the 5-form and the 7-form respectively), which from the field theory point of view would stand for the non-gauge invariance of the operators ${\rm det} A$ and $\epsilon$. On the other hand, the boundary conditions allowing for the D6, D4 would be dual to a certain $SU(N)$ version of the theory, in which the $B$ field would presumably play an important role. Nonetheless, at this point this is no more than a speculation. In particular, the role of the higher curvature couplings, naively coupling the D4 to the 1-form potential (\ref{D0_in_D4}) and thus endowing it with magnetic charge at the same time, remains to be clarified. It should be pointed out that recently a detailed analysis of the field theory has been performed in \cite{Berenstein:2009sa}. Careful analysis of the quantization condition of the $U(1)$ gauge fields suggests that the moduli space of the $U(N)\times U(N)$ gauge theory is a $\mathbb{Z}_k$ cover of the a priori expected ${\rm Sym}_N\,\mathbb{C}^4/\mathbb{Z}_k$, thus allowing for determinant-like operators to be gauge invariant \cite{Berenstein:2009sa} (see also \cite{BT}). These operators are naturally dual to the wrapped D4, which suggests that the $B$ field is turned on. It would be very interesting to clarify the role of the $B$ field in this context, and figure out whether a connection to the possibility raised above, namely the subtle role of the quantization of abelian fields in $AdS$, is possible. Further studies of these issues are well beyond the scope of this paper, and are postponed for further work. 

In this paper we will simply assume that suitable boundary conditions are chosen allowing for the corresponding wrapped objects, and, as we have done for the D4-brane, we will include the effect of the (flat) $B$-field.\footnote{The results for a vanishing $B$-field are simply obtained by tuning the extra worldvolume flux.}  
The D4-brane with zero flux would be identified with the di-baryon operator ${\rm det}\,A=\epsilon_{i_1\dots i_N}\epsilon^{j_1\dots j_N}A^{i_1}_{j_1}\dots A^{i_N}_{j_N}$ in the CFT side, being $A$ one of the bifundamentals in the field theory. It is also natural to ask what could be the dual of the D4-brane with non-minimal flux.
Since once the worldvolume flux is turned on extra F1 strings are required, we should expect such dual operator to involve $n_f=\frac{k\,\mathcal{N}}{2}$ Wilson lines in the fundamental representation of $U(N)\times U(N)$. Indeed, the configuration is reminiscent of the D5 Wilson loop in $AdS_5\times S^5$ \cite{Yamaguchi:2006tq}, which suggests that these fundamental indices should be antisymmetrized.  We will see in the next section that dynamically a bound $n_f^{max}\sim \sqrt{N\, k}\sim \lambda^{-\frac{1}{2}}\, N$, where $\lambda=N/k$ is the 't Hooft coupling, in the number of such fundamental indices appears, which is consistent with the antisymmetrization assumption. It would be interesting to elaborate further on this proposal, and in particular to understand the dependence on the 't Hooft coupling. We postpone such analysis for further work.

\subsection{The 't Hooft monopole}

Let us now consider the D2-brane wrapping the $\mathbb{P}^1$ in $\mathbb{P}^3$, identified in \cite{ABJM} with a ('t Hooft) monopole operator \cite{GNO,B,K}. 

Since this brane captures the $F_2$ flux it requires fundamental strings in order to cancel the worldvolume tadpole. Substituting (\ref{fluxes}) in the CS coupling
\begin{equation}
S_{CS}=2\pi\, T_2\int_{R\times \mathbb{P}^1}P[F_2]\wedge A= k\, T_{F1}\int dt A_t
\end{equation}
we find that the number of fundamental strings must be $q=k$. Note in particular that this is the anticipated result from the di-baryon case, where the tadpole of a single D2 was expected to be $k$.

We are now interested in adding worldvolume flux to this configuration. According to the observation in the previous section, there is a background $B_2$ field  given by (\ref{shift_B2}) \cite{AHHO}. It is then convenient to split the worldvolume flux as in the previous section $F_T=F+J$, with
\begin{equation}
\label{mflux}
F={\cal N} J\, .
\end{equation}
We should stress that the D2-brane, wrapping a spin manifold, does not capture the Freed Witten anomaly, and as such, the quantization condition for $F_T$ \cite{Bachas:2000ik} is

\begin{equation}
\frac{1}{2\pi}\int F_T=\frac{1}{2}\,(\mathcal{N}+1)\in \mathbb{Z}
\end{equation}
Therefore, the case with minimal magnetic flux corresponds to $\mathcal{N}=-1$.

The D2-brane DBI action then reads
\begin{equation}
S_{DBI}=-\frac{\pi\,T_2}{g_s}\, \sqrt{L^4+(2\pi \mathcal{N})^2}\, \int dt\, \frac{2\rho}{L}\, .
\end{equation}
Besides, there is D0-brane charge induced in the configuration, since
\begin{equation}
\label{D0_in_D2}
S_{CS}=2\pi\, T_2\int_{\mathbb{R}\times \mathbb{P}^1} C_1\wedge F=\frac{{\cal N}}{2}\, T_0 \int_{\mathbb{R}} C_1\, .
\end{equation}
Note that even in the case of minimal magnetic flux, ${\cal N}=-1$, there is a non-zero D0-brane charge induced by the shifted $B_2$. 

In this case the charge of the worldvolume tadpole is not modified by the presence of the magnetic flux.

In the next section we will study the dynamics of the configuration formed by the D2-brane plus the $k$ fundamental strings, and show that adding magnetic flux allows to construct more general 't Hooft monopole configurations with charge. This charge will have to lie however below some upper bound for the configuration to be stable in the AdS direction. 

In view of (\ref{D0_in_D2}) we see that our system is actually formed by a $D2-D0$ bound state, which hints to a non-supersymmetric configuration.\footnote{For this reason conjecturing a dual operator seems much harder.} Thus, one might worry about the stability of the configuration with flux. Nevertheless, since both the cycle wrapped by the brane and the worldvolume gauge field are topologically non-trivial, we expect the configuration to be stable, at least under small perturbations. As discussed in the previous subsection, it is implicit in our probe brane approximation that the strings are uniformly distributed over the D2 worldvolume. Grouping them together in a point would require to consider their backreaction on the D2, which would deform it into a spike, which could in turn be unstable due to the lack of SUSY. Nevertheless, as long as we restrict to the probe approximation, we expect the system to be perturbatively stable.

\subsection{The baryon vertex}

Let us finally consider the D6-brane wrapping the whole $\mathbb{P}^3$. This brane is the analogue of the baryon vertex in $AdS_5 \times S^5$ \cite{Witten}. In the absence of worldvolume magnetic flux this brane captures the ${F}_6$ background flux, and it requires the addition of $q=N$ fundamental strings:
\begin{equation}
\label{F1_no_flux}
S_{CS}=2\pi \, T_6\int_{R\times \mathbb{P}^3}P[F_6]\wedge A= N\, T_{F1}\int dt A_t\, .
\end{equation}

Note however that once the shift in (\ref{shift_B2}) has been taken into account, the above expression is incomplete, since there are extra contributions to the F1 charge coming from the coupling $\int F_2\wedge B_2\wedge B_2$. Nevertheless, once the higher curvature corrections are taken into account, they cancel out so that the correct expression is actually (\ref{F1_no_flux}). 
 In the case at hand the relevant term in (\ref{Higher_Curv}) is 
\begin{equation}
\label{Higher_Curvature_In_This_Case}
\Delta S=\frac32 (2\pi)^5\, T_6\int C_1\wedge F\wedge \frac{1}{48}\, (\hat{p}_1(N)-\hat{p}_1(T))
\end{equation}
As shown in \cite{AHHO}  this term contributes to the D6-brane action inducing extra F1 charge as
\begin{equation}
\Delta S=-\frac18 (2\pi)^6\, k \,T_6 \,\int dt\, A_t\, ,
\end{equation}
and this precisely cancels the $B_2$ contribution to (\ref{F1_no_flux}).

Let us now switch on a gauge flux, $F_T={\cal N}J$. Note that this represents a slight change in the conventions compared to the previous sections, where we split $F_T$ into two pieces one cancelling $B$. In this case, due to the relevance of the curvature coupling in giving the tadpole of $N$ units in the unfluxed case, it turns out to be more convenient not to do the spliting so that the argument as in \cite{AHHO} goes through. Since $\mathbb{P}^3$ is spin, the appropriate quantization condition is
\begin{equation}
\label{quant}
\frac{1}{2\pi}\int F_T=\frac{{\cal N}}{2} \in \mathbb{Z}
\end{equation}
%
%
%As in the previous subsections, it turns out to be useful to split it as $F_T=F+J$, such that $J$ cancels the $B_2$ shift and we can effectively work as if we had no $B_2$ field and ${\cal F}=2\pi F$ with $F={\cal N}J$. However, as in the D2-brane case, since $\mathbb{P}^3$ is spin, the appropriate quantization condition will be
%\begin{equation}
%\frac{1}{2\pi}\int F_T=\frac{1}{2}\, (\mathcal{N}+1)\in \mathbb{Z}
%\end{equation}
%

The DBI action of the D6-brane becomes:
\begin{equation}
S_{D6}=-\frac{\pi^3\, T_6}{6\, g_s}\, \Bigl(L^4+(2\pi\,(\mathcal{N}-1))^2\Bigr)^{3/2}\,  \int dt\, \frac{2\, \rho}{L}\, .
\end{equation}

In this case the magnetic flux modifies the number of fundamental strings that must end on the D6, since it contributes to the worldvolume tadpole through the couplings
\begin{equation}
S_{CS}=\frac16 (2\pi)^2\, T_6 \int_{\mathbb{R}\times \mathbb{P}^3} P[F_2]\wedge F_T \wedge \Bigl(2\pi F_T +3\, P[B_2]\Bigr)\wedge A=
k\, \frac{{\cal N}({\cal N}-2)}{8}\, T_{F1} \int dt A_t\, .
\end{equation} 
Therefore  $q=N+k\,{\cal N}({\cal N}-2)/8$. Note that this is always an integer for the quantization
condition (\ref{quant}). As for the D4-brane, this is the number of fundamental strings required to cancel the tadpole of each of the D2-branes that are dissolved on the D6-brane by the magnetic flux and the $B_2$ field, through the coupling:
\begin{equation}
S_{CS}=\frac12 \, T_6\int_{\mathbb{R}\times \mathbb{P}^3} C_3\wedge {\cal F}\wedge {\cal F}
\end{equation}
In this coupling the term proportional to $\int C_3\wedge B_2\wedge B_2$ is precisely cancelled with the contribution of the A-roof $\int C_3\wedge \frac{1}{48}\, (\hat{p}_1(N)-\hat{p}_1(T))$. The other two terms give
\begin{equation}
S_{CS}=\frac12 (2\pi)^2\, T_6\int C_3 \wedge F_T \wedge F_T + (2\pi) \, T_6 \int C_3 \wedge F_T \wedge B_2=\frac{{\cal N}({\cal N}-2)}{8}\, T_2 \int C_3
\end{equation}

Note that the magnetic flux and the $B_2$ field also induce D0-brane charge in the configuration.

%Note that given that $F$ is proportional to the K\"ahler form on the $\mathbb{P}^3$ it satisfies that  $\int_{\mathbb{P}^3}F\wedge F\wedge F={\cal N}^3 \pi^3$, and therefore it also induces D0-brane charge in the configuration 
%\begin{equation}
%S_{CS}=\frac16 (2\pi)^3\, T_6 \int_{\mathbb{R}\times \mathbb{P}^3} C_1 \wedge F \wedge F \wedge F =\frac{{\cal N}^3}{48}\, T_0 \int C_1\, .
%\end{equation}
%
%However, from the higher curvature coupling (\ref{Higher_Curvature_In_This_Case}) there is yet another contribution
%\begin{equation}
%\Delta S= -\frac{{\cal N}}{192}\, T_0 \int C_1\
%\end{equation}
%
%so that the total D0 charge is
%\begin{equation}
%\frac{{\cal N}}{48}({\cal N}^2-\frac14)\, T_0 \int C_1\, .
%\end{equation}
%

We will study the dynamics of the D6-brane with magnetic flux in the next section. We will see that, similarly to the D2-brane case, adding magnetic flux allows to construct more general baryon vertex configurations in which the charge of the brane can be increased up to some maximum value. In this case, since the number of fundamental strings attached to the D6-brane depends on the magnetic flux, the bound on the magnetic flux imposes as well a bound on the number of F-strings that can end on the brane.

As in the D2 brane case, induced D0 brane charge  in a D6 suggests that the system will not be supersymmetric. However, again due to its non-trivial topology, we expect the system to be perturbatively stable.

\section{Study of the dynamics: Charge bounds}

In this section we study the stability in the $AdS$ direction of the brane configurations that we have previously discussed. We follow the calculations in \cite{BISY} and \cite{Maldacena} (see also \cite{JLR} for similar results for the baryon vertex with magnetic flux in $AdS_5\times S^5$). We show that the energy of the various configurations is inversely proportional to the distance between the quarks, as predicted by conformal invariance, and that the proportionality constant is negative, so that the configurations are stable against perturbations in $\rho$.
As expected, we find the same non-analytical behavior with the square root of the 't Hooft coupling that was found for the baryon vertex in $AdS_5\times S^5$ \cite{BISY} and the $q{\bar q}$ system  \cite{Maldacena,RY}. This represents a non-trivial prediction of AdS/CFT for the strongly coupled CS-matter theory. 

In order to analyze the stability in the $\rho$-direction we have to consider both the D$p$-brane wrapped on the $\mathbb{P}^{p/2}$ and the $q$ fundamental strings stretching between the brane and the boundary of $AdS$. The action is given by
\begin{equation}
S=S_{Dp}+S_{qF1}\, ,
\end{equation}
where $S_{Dp}$ is of the form\footnote{Note that for the D6-brane ${\cal N}\rightarrow {\cal N}-1$ in $Q_p$ in order to account for the $B_2$ field, consistently with the quantization condition (\ref{quant}). We will take due care of this shift in section 3.3 below.}
\begin{equation}
S_{Dp}=-Q_p\int dt \, \frac{2\rho}{L}\, , \qquad {\rm with}\qquad  Q_p=\frac{\pi^{p/2}\,T_p}{(\frac{p}{2})!\, g_s}(L^4 + (2\pi{\cal N})^2)^{p/4}\, , 
\end{equation}
and the action of the strings is given by
\begin{equation}
S_{qF_1}=-q\,T_{F1}\int dt dx\,\sqrt{\frac{16 \rho^4}{L^4}+\rho'^2}\, ,
\end{equation}
where we have parameterized the worldvolume coordinates by $(t,x)$ and the position in $AdS$ by $\rho=\rho(x)$. Following the analysis in \cite{BISY} the equations of motion come in two sets: the bulk equation of motion for the strings, and the boundary equation of motion (as we are dealing with open strings), which contains as well a term coming from the D$p$-brane. One can show easily that these equations of motion are, respectively:
\begin{equation}
\frac{\rho^4}{\sqrt{\frac{16\rho^4}{L^4}+\rho'^2}}=c
\end{equation}
with $c$ some constant, and
\begin{equation}
\label{boundary}
\frac{\rho_0'}{\sqrt{\frac{16\rho_0^4}{L^4}+\rho_0'^2}}=\frac{2Q_p}{L\,q\,T_{F_1}}\, ,
\end{equation}
where $\rho_0$ is the position of the brane in the holographic direction and $\rho^\prime_0=\rho^\prime(\rho_0)$. As in \cite{BISY,JLR} it is convenient to define
\begin{equation}
\label{beta}
\sqrt{1-\beta^2}=\frac{2 Q_p}{L\,q\,T_{F_1}}\, ,
\end{equation}
where $\beta\in [0,1]$. The two equations of motion can then be combined into just
\begin{equation}
\label{eqmotion}
\frac{\rho^4}{\sqrt{\frac{16\rho^4}{L^4}+\rho'^2}}=\frac{1}{4}\,\beta\, \rho_0^2\, L^2\, .
\end{equation}

Integrating the equation of motion we find that the size of the configuration is given by
\begin{equation}
\label{size}
{\ell}=\frac{L^2}{4\rho_0}\int_1^\infty dz \frac{\beta}{z^2\sqrt{z^4-\beta^2}}\, ,
\end{equation}
where $z=\rho/\rho_0$. This expression has the same form as the size of the baryon vertex in $AdS_5 \times S^5$ \cite{BISY}\footnote{And also that of the baryon vertex with magnetic flux constructed in \cite{JLR}.} and the $q{\bar q}$ system \cite{Maldacena,RY}, and can also be solved in terms of hypergeometric functions. Note that the dependence on the location of the configuration, $\rho_0$, and on $L^2$ is also the same, which is again a non-trivial prediction of the AdS/CFT correspondence for the strong coupling behavior of the gauge theory.

The total on-shell energy is given by
\begin{eqnarray}
E&=& E_{Dp} + E_{qF1} =q\, T_{F_1}\,\rho_0\Big(\frac{2Q_p}{L\,q\,T_{F_1}}+\int_1^{\infty}dz\frac{z^2}{\sqrt{z^4-\beta^2}}\Big) \nonumber\\
&=& q T_{F_1}\rho_0\Big(\sqrt{1-\beta^2}+\int_1^{\infty}dz\frac{z^2}{\sqrt{z^4-\beta^2}}\Big)\, .
\end{eqnarray}

The binding energy of the configuration can be obtained by subtracting the (divergent) energy of its constituents. When the D$p$-brane is located at $\rho_0=0$ the strings stretched between 0 and $\infty$ become radial, and therefore correspond to free quarks. At this location the energy of the D$p$ vanishes. Therefore, the binding energy is given by:
\begin{equation}
\label{Ebinding}
E_{\rm bin}= q\, T_{F_1}\,\rho_0\Big(\sqrt{1-\beta^2}+\int_1^{\infty}dz\Big[\frac{z^2}{\sqrt{z^4-\beta^2}}-1\Big]-1\Big)\, .
\end{equation}
This expression has again the same form than the corresponding expressions in \cite{BISY,JLR,Maldacena,RY}.\footnote{In this case we have added the on-shell energy of the D$p$-brane.}

Notice that for our configurations $\beta$ is a function of the magnetic flux that is dissolved on the D$p$-brane, since from  (\ref{beta})
\begin{equation}
\label{beta2}
\beta=\sqrt{1-\Bigl(\frac{2Q_p}{L\,q\,T_{F1}}\Bigr)^2}\, .
\end{equation}
In particular, in order to find a stable configuration we must have
\begin{equation}
\label{bound}
\frac{2 Q_p}{L\,q\,T_{F_1}}\le 1\, .
\end{equation}
This imposes a bound on the possible values of the magnetic flux, and therefore on the possible charges that can be dissolved in the D$p$-brane. This situation is very similar to the one found in \cite{JLR} for the baryon vertex in $AdS_5\times S^5$ with magnetic flux.
Moreover, for the di-baryon and baryon vertex configurations, for which the number of fundamental strings required to cancel the tadpole depends on the magnetic flux, there is as well a bound on the number of quarks that can form the bound state.

For the values allowed by (\ref{bound}) the binding energy per string is negative and decreases monotonically with $\beta$. \footnote{Its behavior as a function of the magnetic flux depends on the specific function $\beta({\cal N})$ given by (\ref{beta2}) We will analyze this behavior in the next subsections for the different branes.}. Therefore, the configuration is stable, becoming less and less stable as $\beta$ decreases, with the binding energy reaching its maximum value at  the bound, $\beta=0$, where it vanishes. The configuration reduces then to $q$ free radial strings stretching from $\rho_0$ to $\infty$, plus a D$p$-brane located at $\rho_0$. Note that this configuration only exists when the magnetic flux is non-vanishing, since only then we can reach $\beta=0$. When the D$p$-brane is charged the configuration corresponding to free quarks is therefore degenerate. It can be realized either as free radial strings stretching from 0 to $\infty$ plus a charged D$p$-brane, with the charge satisfying (\ref{bound}), located at $\rho=0$, or as free radial strings stretching from  $\rho_0$ to $\infty$ plus a D$p$-brane located at $\rho_0$, with a charge that has to satisfy the equality in (\ref{bound}). In this case the F1's are less energetic due to the fact that they now stretch from $\rho_0$ to $\infty$ but this is compensated by the energy of the brane at $\rho_0$, charged such that $\beta=0$. Note that the location of the D$p$-brane has become a moduli of the system.
In both cases since the strings are radial the size of the configuration vanishes.

Note that from (\ref{Ebinding}) and (\ref{size}) we have that for all values of the magnetic charge
\begin{equation}
\label{stability}
E_{\rm bin}=-f(\beta) \frac{(g_s N)^{2/5}}{\ell}
\end{equation}
with $f(\beta)\ge 0$. Therefore $dE/d{\ell}\ge 0$ and the configuration is stable. The behavior of $E_{\rm bin}$ as a function of the 't Hooft coupling and the size of the configuration is the same as in $AdS_5\times S^5$ \cite{Maldacena,RY,BISY}. As in that case the fact that it goes as $1/{\ell}$ is dictated by conformal invariance, while the behavior with $\sqrt{\lambda}$ is a non-trivial prediction for the non-perturbative regime of the gauge theory. Note that the same non-analytic behavior with $\lambda$ is predicted for ${\cal N}=4$ SYM in 3+1 dimensions and for ABJM \cite{DPY,CW,RSY}. In fact, perturbative calculations like those in \cite{ESZ,DG,Pestun} can explain this behavior when extrapolated to strong coupling, as inferred in \cite{Suyama}. Further, the exact interpolating function between the weak and strong coupling regimes for 1/6 and 1/2 BPS Wilson loops was obtained in \cite{MP}, using topological strings and the localization techniques applied in \cite{KWY} to ABJM theories. \footnote{See also \cite{DT}.}
 
We have plotted in Figure 1 the behavior of $f(\beta)/qT_{F1}$ as a function of $\beta$. We can see that when the number of strings does not depend on $\beta$, i.e. for the 't Hooft monopole case, the configuration becomes more stable as $\beta$ increases. For the di-baryon and baryon vertex configurations the number of strings changes with the magnetic flux, and therefore the stability of the configuration will vary with $\beta$ in a way which depends on the specific function (\ref{beta2}). We will analyze this behavior in the next subsections. 

\begin{figure}[t]
\begin{center}
\includegraphics[scale=.8]{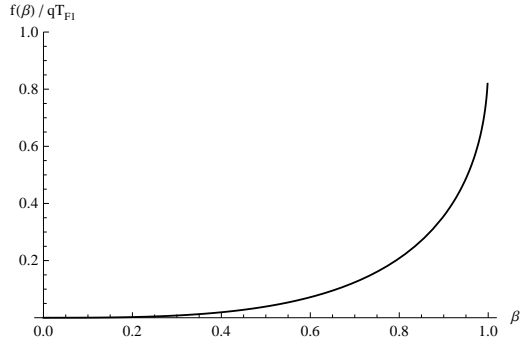}
\caption{Stability of the configuration, for fixed number of strings, as a function of $\beta$.}  
\end{center}
\label{fig:fbeta}
\end{figure}

\vspace{0.3cm}

We now discuss in some more detail the dynamics of the different configurations discussed in the previous section.

\subsection{The 't Hooft monopole}

In this case 
\begin{equation}
Q_2=\frac{\pi\, T_2}{g_s}\sqrt{L^4 + (2\pi{\cal N})^2}
\end{equation}
and $q=k$, so that 
\begin{equation}
\beta=\sqrt{1-\frac{1}{4\pi^2}\Bigl(1+\frac{4\pi^2 {\cal N}^2}{L^4}\Bigr)}\, .
\end{equation}
The behavior of the binding energy as a function of ${\cal N}$ is shown in Figure 2.
The minimum binding energy occurs for zero ${\cal N}$, for which  $\beta=\sqrt{1-\frac{1}{4\pi^2}}$, and $\beta=0$ is reached for
\begin{equation}
\label{firstbound}
\frac{{\cal N}_{\rm max}}{L^2}=  \sqrt{1-\frac{1}{4\pi^2}}\, ,
\end{equation}
for which the monopole is no longer stable and reduces to $k$ radial F1's, stretching from $\rho_0$ to $\infty$, plus a spherical D2-brane with D0-charge 
$\frac{L^2}{2}\sqrt{1-\frac{1}{4\pi^2}}$,
located at $\rho_0$. As a function of the 't Hooft coupling (\ref{firstbound}) becomes
\begin{equation}
\label{firstbound2}
{\cal N}_{\rm max}=\sqrt{8\lambda\,(4\pi^2-1)}
\end{equation}
which is exactly the same behavior that was encountered in \cite{JLR} for the maximum value of the magnetic flux that could be dissolved in the baryon vertex in $AdS_5\times S^5$. We will find this same behavior for the di-baryon and baryon vertex with flux in the next subsections. 
Although dynamically the origin of the bound is quite clear, pointing at an instability when the magnetic flux makes the energy of the brane too large, its interpretation from the CFT side is not clear to us. We refer to the conclusions for a brief discussion.

\begin{figure}[t]
\begin{center}
\includegraphics[scale=.8]{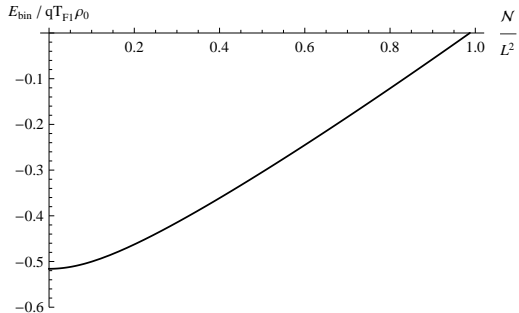}
\caption{Binding energy per string of the 't Hooft monopole (in units of $T_{F1}\, \rho_0$)  as a function
 of ${\cal N}/L^2$.}  
\end{center}
\label{fig:EnergyD2perstring}
\end{figure}

\subsection{Di-baryon vertex}

In this case
\begin{equation}
Q_4=\frac{\pi^2\, T_4}{2 g_s}\Bigl(L^4+ (2\pi{\cal N})^2\Bigr)
\end{equation}
and $q=k\,{\cal N}/2$, so that      
\begin{equation}
\label{betaD4}
\beta=\sqrt{1-\frac{L^4}{64\pi^4{\cal N}^2}\Bigl(1+ \frac{4\pi^2 {\cal N}^2}{L^4}\Bigr)^2}\, .
\end{equation}
This function has a maximum at $\frac{{\cal N}}{L^2} = \frac{1}{2\pi}$, where it reaches 
$\beta_{max}=\sqrt{1-\frac{1}{4\pi^2}}$. For this value of the magnetic flux the binding energy per string is minimum. Note however that since the number of strings depends also on ${\cal N}$ this is not the value for which the configuration is maximally stable (if we define the stability in terms of the function $f(\beta)$ in (\ref{stability})).
The allowed values for the magnetic flux are those for which $\beta \in [0,\beta_{max}]$:
\begin{equation}
\label{intervalD4}
 1-\sqrt{1-\frac{1}{4\pi^2}}\le \frac{{\cal N}}{L^2}\le 1+\sqrt{1-\frac{1}{4\pi^2}}\, .
\end{equation}
At both ends $\frac{{\cal N}_\pm}{L^2}=1\pm \sqrt{1-\frac{1}{4\pi^2}}$, $\beta=0$, and the configuration turns into a collection of
$q=k\,{\cal N}_\pm/2  $ free quarks plus a wrapped D4-brane. The behavior of the binding energy per string as a function of ${\cal N}/L^2$ is shown in Figure 3 (left). Since the total binding energy of the configuration depends on the number of strings, which is a function of the magnetic flux, the behavior of the binding energy is modified as shown in Figure 3 (right). The minimum energy occurs now for ${\cal N}=1.00 \, L^2$.
 In Figure 4 we have plotted as well $f(\beta)/T_{F1}$ (see (\ref{stability})) as a function of the magnetic flux.

\begin{figure}[t]
  \begin{center}
  $\begin{array}{c@{\hspace{0.6in}}c}
    \includegraphics[scale=.8]{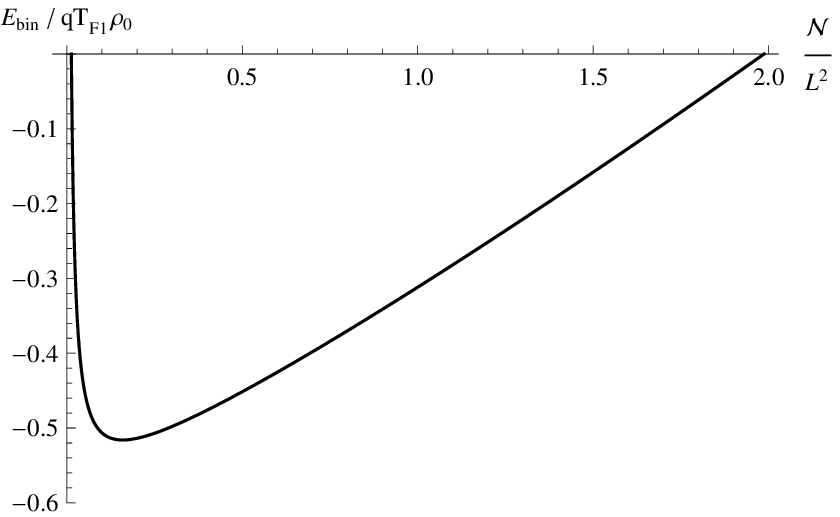}
    & \includegraphics[scale=.8]{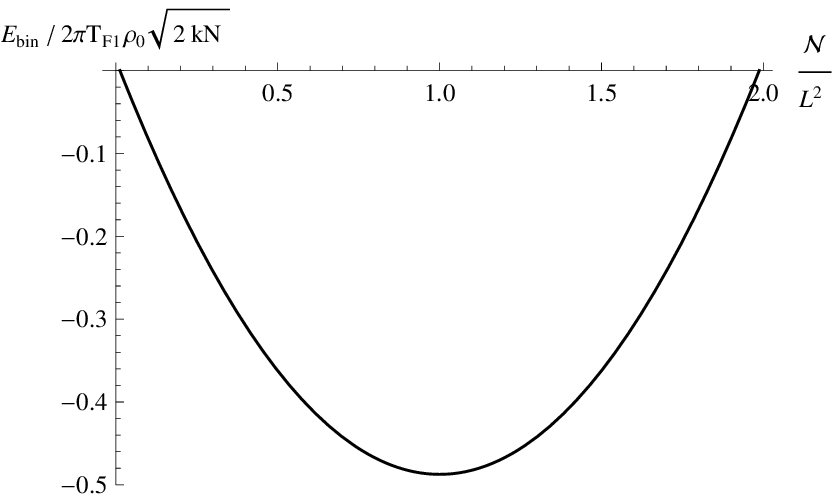}
  \end{array}$
  \end{center}
  \caption{Binding energy per string (left) and total binding energy (right) of the di-baryon (in units of $T_{F1}\, \rho_0$ and $2\pi\, T_{F1}\, \rho_0\, \sqrt{2kN}$ respectively)  as a function
 of ${\cal N}/L^2$.}
  \label{fig:EnergyD4}
\end{figure}

\begin{figure}[t]
\begin{center}
\includegraphics[scale=.8]{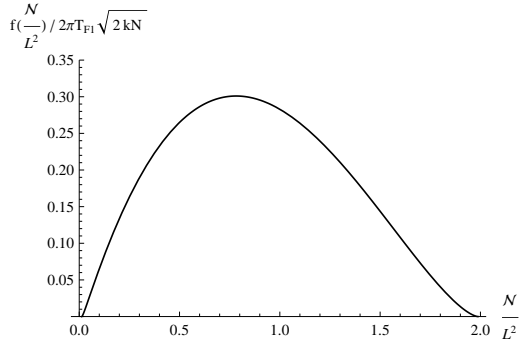}
\caption{$f(\beta)$ for the di-baryon, in units of $2\pi\, T_{F1}\, \sqrt{2kN}$, as a function
 of the magnetic flux.}  
\end{center}
\label{fig:fbetaD4}
\end{figure}

As we have seen, the D4-brane with flux exhibits a very different behavior with the magnetic flux than the D2-brane.\footnote{And the D6-brane, as we will see next.}
 The main difference is coming from the fact that now the magnetic flux induces a worldvolume tadpole in the D4-brane that was not present for ${\cal N}=0$, and this tadpole has to be cancelled by adding a number of F1's proportional to ${\cal N}$. Accordingly, the whole configuration of point-like D4-brane plus fundamental strings only exists for ${\cal N}\ne 0$, with
the allowed interval for the magnetic flux, given by (\ref{intervalD4}), implying an allowed interval for the number of fundamental strings ending on the D4-brane:
\begin{equation}
\label{interval}
2\pi\, \sqrt{2kN}\Bigl( 1-\sqrt{1-\frac{1}{4\pi^2}}\Bigr)\le q \le 2\pi\, \sqrt{2kN} \Bigl( 1+\sqrt{1-\frac{1}{4\pi^2}}\Bigr)\, .
\end{equation}
At the bounds the strings become radial and the configuration ceases to be stable.

\subsection{Baryon vertex}

In this case
\begin{equation}
Q_6=\frac{\pi^3\, T_6}{6g_s}\Bigl(L^4+(2\pi ({\cal N}-1))^2\Bigr)^{3/2}
\end{equation}
and $q=N+k\,{\cal N}({\cal N}-2)/8$, so that
\begin{equation}
\label{betaD6}
\beta=\sqrt{1-\frac{1}{36\pi^2\Bigl(1+\frac{4\pi^2 {\cal N}({\cal N}-2)}{L^4}\Bigr)^2}\Bigl(1+\frac{4\pi^2({\cal N}-1)^2}{L^4}\Bigr)^3}\, .
\end{equation}
This function decreases monotonically with ${\cal N}$, reaching its minimum value $\beta=0$ when $\frac{{\cal N}}{L^2}\sim \frac{\sqrt{36\pi^2-1}}{2\pi}$. 
Therefore the allowed values of the magnetic flux are
\begin{equation}
\label{intervalD6}
\frac{{\cal N}}{L^2}\lesssim \frac{\sqrt{36\pi^2-1}}{2\pi}
\end{equation}
We have plotted in Figure 5 (left) the binding energy per string as a function of ${\cal N}/L^2$. We can see that the qualitative behavior 
is very similar to the D2-brane case, and also to the charged baryon vertex in $AdS_5 \times S^5$ \cite{JLR}. In all these examples the binding energy per string increases with the magnetic flux till it becomes zero when the strings are radial and the baryon size vanishes. Note however that in this case the tadpole induced in the worldvolume of the D6-brane depends on the magnetic flux, and therefore the number of quarks that can form the bound state depends on ${\cal N}$, as $q=N+k\,{\cal N}({\cal N}-2)/8$. This modifies the behavior of the total binding energy as shown in Figure 5 (right).
Here we can see that the minimum energy configuration occurs for ${\cal N}/L^2 \sim 2.01$, and that the configuration loses stability till it reduces to free radial fundamental strings at
${\cal N}_{max}/L^2\sim \frac{\sqrt{36\pi^2-1}}{2\pi}$, for which  $q\sim 36\pi^2 N$. The stability of the configuration as a function of the magnetic flux can be seen in Figure 6.

\begin{figure}[t]
  \begin{center}
  $\begin{array}{c@{\hspace{0.6in}}c}
    \includegraphics[scale=.8]{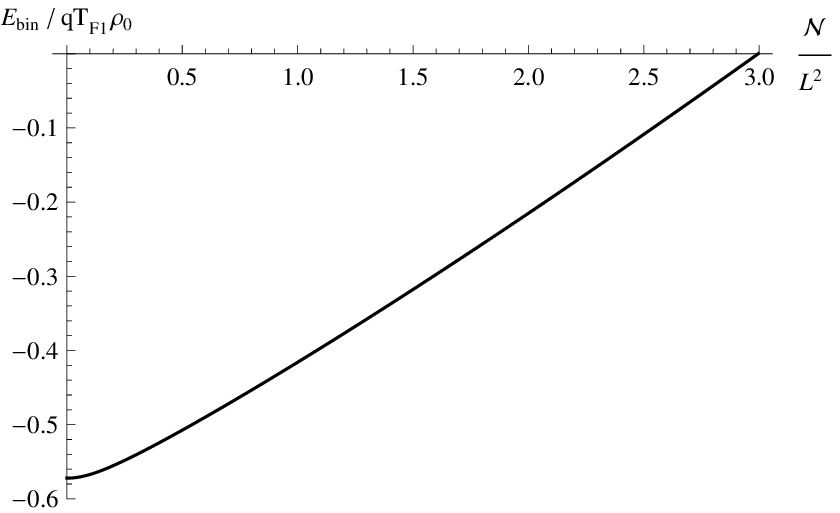}
    & \includegraphics[scale=.8]{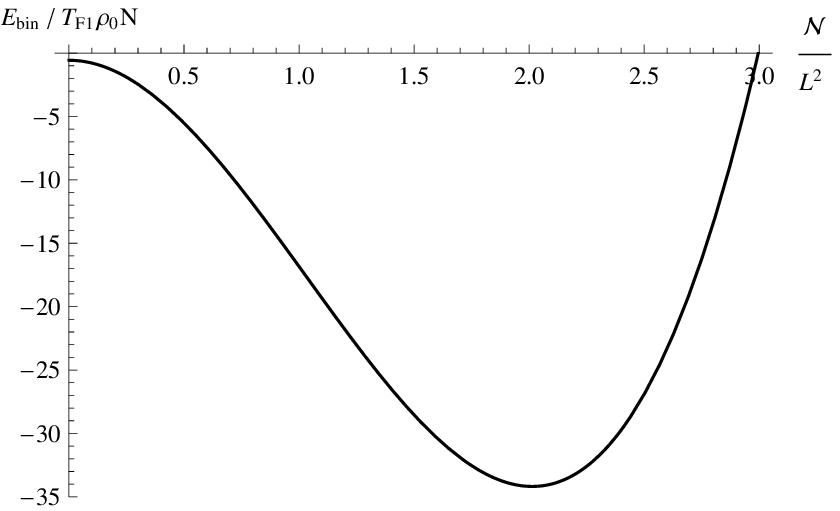}
  \end{array}$
  \end{center}
  \caption{Binding energy per string (left) and total binding energy (right) of the baryon vertex (in units of $T_{F1}\, \rho_0$ and $T_{F1}\, \rho_0\, N$ respectively)  as a function
 of ${\cal N}/L^2$.}
  \label{fig:EnergyD6}
\end{figure}

\begin{figure}[t]
\begin{center}
\includegraphics[scale=.8]{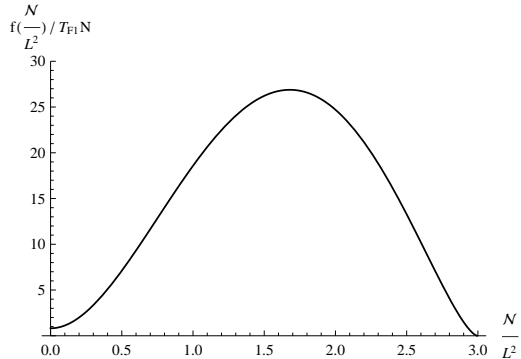}
\caption{$f(\beta)$ for the baryon vertex, in units of $T_{F1}\, N$,  as a function
 of the magnetic flux.}  
\end{center}
\label{fig:fbetaD6}
\end{figure}

The analysis in this section shows that the addition of magnetic flux to the D6-brane allows the construction of more general baryon vertex configurations in  which the number of quarks can be increased up to $\sim 36\pi^2 N$. A way to construct baryons with $q<N$ number of quarks in $AdS_5 \times S^5$ was considered in \cite{BISY}. In this background $q=N$ strings are needed in order to cancel the tadpole in the worldvolume of the spherical D5-brane, $N$ being the rank of the gauge group.  It is however possible to find more general baryon vertex configurations with $q<N$ quarks if $N-q$ strings stretch between $\rho_0$ and 0. The study of the dynamics of these configurations reveals that they are stable if the number of quarks satisfies
$5N/8\le q \le N$. For the minimum value, $q_{min}=5N/8$, the strings are radial and the binding energy vanishes, exactly the same behavior that we have found at the limiting values.

A similar analysis to the one in \cite{BISY} for the D6-brane wrapped in $\mathbb{P}^3$ shows that
the boundary equation (\ref{boundary}) has to be modified as
\begin{equation}
\frac{\rho_0'}{\sqrt{\frac{16\rho_0^4}{L^4}+\rho_0'^2}}=\frac{N}{6\pi q}
\Bigl(1+\frac{4\pi^2({\cal N}-1)^2}{L^4}\Bigr)^{3/2}+\frac{1}{q}\Bigl(N+\frac{k {\cal N}({\cal N}-2)}{8}-q\Bigr)\, ,
\end{equation}
from which we can conclude that the number of quarks has to satisfy:
\begin{equation}
\frac12 (N+\frac{k{\cal N}({\cal N}-2)}{8})(1+\sqrt{1-\beta^2})\le q \le N+\frac{k{\cal N}({\cal N}-2)}{8}\, ,
\end{equation}
with $\beta$ given by (\ref{betaD6}). 

Therefore we have found that  by combining the addition of magnetic flux and the construction in \cite{BISY} it is possible to find more general baryon vertex configurations in which the number of quarks differs from $N$ in a way that depends on the magnetic flux dissolved in the D6-brane and the number of strings that end on 0 instead of $\infty$. Like for all the bounds found in this paper, the quarks are free for the minimum and maximum numbers allowed, where the configuration ceases to be stable.

\section{Adding Romans mass}

In this section we briefly discuss how the results of the previous sections for the 't Hooft monopole, di-baryon and baryon vertex configurations are modified by the presence of a non-zero Romans mass $F_0$. 

It was shown in \cite{Gaiotto:2009mv} that the CS-matter theory dual to a perturbation of the previous $AdS_4\times \mathbb{P}^3$ background by a mass term should be a perturbation of ABJM with levels $k_1+k_2=F_0$. The simplest way to see this is to realize that a D0-brane in this background develops a tadpole through its CS coupling to the mass \cite{GHT}:
\begin{equation}
S_{CS}=T_0\int dt\,  F_0\, A_t\, ,
\end{equation}
and therefore $F_0$ fundamental strings should end on it. One can account for these extra indices in the fundamental by modifying the level of one of the gauge groups, such that the di-monopole operator dual to the D0-brane becomes
\begin{equation}
O^{D0}=({\rm Sym_k})_{i_1\dots i_{k+F_0}}(\overline{{\rm Sym}_k})^{j_1\dots j_k}A^{i_1}_{j_1}\dots A^{i_k}_{j_k}\, .
\end{equation}

It was shown in \cite{Gaiotto:2009mv} that indeed ABJM can be deformed in different ways such that the levels do not sum to zero. In all cases the deformed theory flows to a CFT, with different amounts of supersymmetries and global symmetries.
The theory that is obtained from ABJM by simply changing the CS levels such that $k_1+k_2$ is small (in the precise way shown in \cite{Gaiotto:2009mv}) breaks all the supersymmetries, but flows to a CFT that respects the SO(6) R-symmetry. This is the theory that can be most simply identified as a deformation of the ${\cal N}=6$ solution by a Romans mass, and the one that we will consider in this section.

The gravity dual of the ${\cal N}=0$ CFT with SO(6) global symmetry discussed in \cite{Gaiotto:2009mv} can be constructed as a perturbation of the ${\cal N}=6$ solution, with the usual Fubini-Study metric on $\mathbb{P}^3$, by a small mass $F_0<< k,N$. In that case the $F_2$ and $F_6$ fluxes are not modified, and the $F_4$ flux that has to be introduced along with the mass (see \cite{Gaiotto:2009mv}) can be compensated with the coupling of $F_2$ with a closed $B_2$ field.
This $B_2$ field will be conveniently absorbed in our definition of $F$. Note however that it contributes to higher order in the mass to expression (\ref{CSmass}). Therefore we will ignore it in our analysis below. Moreover, as in \cite{Gaiotto:2009mv}, we will ignore the effect of the Freed-Witten anomaly. The CS coupling to the mass in the D4-brane case, given by equation (\ref{D4mass}) below, suggests that  a fractional number of F-strings should be added to the D4-brane in order to cancel the tadpole induced by the mass and the Freed-Witten worldvolume flux. Therefore including this effect requires a more careful study, that we hope to address in a future publication.

In this massive background the D0-brane acquires a tadpole. This is however not the case for the other particle-like branes\footnote{If we ignore the effect of the Freed-Witten worldvolume flux, as in \cite{Gaiotto:2009mv}.}, since the only modification in the action 
in the massive $AdS_4\times \mathbb{P}^3$ background is the coupling to the mass \cite{GHT}
\begin{equation}
\label{CSmass}
S_{CS}=T_p \int F_0\, \sum_{r=0}\frac{1}{(r+1)!}(2\pi)^r\, A\wedge F^r\,
\end{equation}
in the CS part.
 
Let us now add a magnetic flux as we did in the previous sections, $F={\cal N}J$. A D2-brane wrapped on $S^2$ will now develop a tadpole proportional to the mass, given that in the Chern-Simons action:
\begin{equation}
S_{CS}=2\pi\, T_2 \int F_0\, A\wedge dA=\frac{F_0\,{\cal  N}}{2}\int dt A_t\, .
\end{equation}
Therefore, for non-zero mass we have to add a number of F1's that is proportional to the product of the mass with the magnetic flux: $q=F_0\, {\cal N}/2$.

For a D4 wrapped on the $\mathbb{P}^2$ the relevant coupling is
\begin{equation}
\label{D4mass}
S_{CS}=\frac{1}{3!} (2\pi)^2\, T_4\int F_0\, A\wedge dA\wedge dA\, ,
\end{equation}
therefore for a non-vanishing magnetic flux we need $q=F_0\,{\cal N}^2/8$ F1's. Note that this is the number of fundamental strings required to cancel the tadpole of the D0-branes dissolved in the D4-brane in the massive case.
Finally, for a D6-brane wrapped on the $\mathbb{P}^3$ the relevant coupling is
\begin{equation}
S_{CS}=\frac{1}{4!}(2\pi)^3\,T_6\int F_0\, A\wedge dA\wedge dA\wedge dA
\end{equation}
and the number of F1's that must be added for non-zero mass is
$q=F_0\,{\cal N}^3/48$, which is again $F_0$ times the number of D0-branes dissolved in the D6-brane.

We have summarized in Table 1 the number of fundamental strings that are required in order to cancel the tadpoles originating from all the background fluxes for each type of wrapped brane.

\begin{table}[!ht]
\begin{center}
\begin{tabular}{||c|c||}
\hline\hline
D$p$-brane & Number of F1's      \\
\hline\hline
D0
& $F_0$ \\
\hline
D2 & $k+\frac{F_0 {\cal N}}{2}$   \\
\hline
D4 & $\frac{k{\cal N}}{2}+\frac{F_0 {\cal N}^2}{8}$ \\ 
\hline
D6 & $N+\frac{k{\cal N}^2}{8}+\frac{F_0{\cal N}^3}{48}$ \\
\hline\hline
\end{tabular}
\end{center}  
\caption{\label{table1} Number of F1's that must end on each D$p$-brane in the presence of mass (and magnetic flux).}
 \end{table}

Note that although $F_0<<k,N$, ${\cal N}$ can be sufficiently large so as to make $F_0{\cal N}\approx k$. This is certainly the case for the values found in (\ref{firstbound}), (\ref{intervalD4}), (\ref{intervalD6}). In the next section we study the dynamics of the particle-like brane configurations with these F1's attached.

\subsection{Dynamics}

The dynamics of the various brane configurations discussed in section 3 is modified in the presence of a non-zero mass due to the fact that the number of F1's attached to the brane depends on the mass as shown in Table 1.

Let us consider first the di-monopole, or D0-brane. Following the analysis in section 3 we have that
$Q_0=T_0/g_s$ and $q=F_0$. Therefore, 
\begin{equation}
\beta=\sqrt{1-\Bigl(\frac{2k}{F_0L^2}\Bigr)^2}
\end{equation}
and the bound (\ref{bound}) leads to 
\begin{equation}
\label{boundmass1}
F_0\ge \frac{2k}{L^2}
\end{equation} 
Therefore, the configuration is stable if the mass is sufficiently large. Note that this bound is perfectly compatible, in the regime of validity of the supergravity description, with the fact that $F_0<<k,N$.
We have plotted in Figure 7 the behavior of the binding energy per string as a function of the mass. Here we see that the configuration is maximally stable when $F_0\rightarrow\infty$, for which $\beta_{\rm max}=1$, and reduces to $F_0$ free quarks plus a D0 at the bound, when $\beta=0$.

\begin{figure}[t]
\begin{center}
\includegraphics[scale=.8]{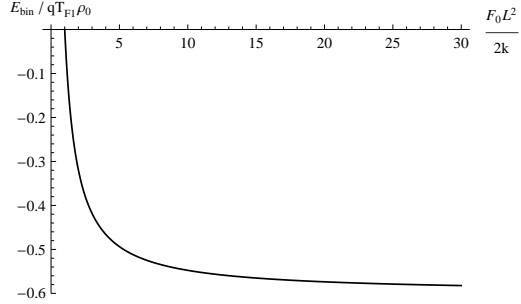}
\caption{Binding energy per string of the di-monopole, in units of $T_{F1}\, \rho_0$,  as a function
 of $\frac{F_0 L^2}{2k}$.}  
\end{center}
\label{fig:EnergyD0perstringwithF0}
\end{figure}

The D2-brane with flux turns out to be more stable in the presence of mass. In this case
\begin{equation}
\beta=\sqrt{1-\frac{1}{4\pi^2(1+\frac{F_0{\cal N}}{2k})^2}\Bigl(1+\frac{4\pi^2{\cal N}^2}{L^4}\Bigr)}
\end{equation}
This function has a maximum at ${\cal N}=\frac{F_0 L^4}{8\pi^2 k}$. Since the binding energy per string decreases monotonically with $\beta$ this is the value of the magnetic flux for which the binding energy (per string) is minimum.
 
The values of the magnetic flux for which the configuration can form a bound state depend on the mass. When $F_0$ satisfies the bound (\ref{boundmass1}), required by the stability of the D0-brane, 
the configuration is stable for all values of the magnetic flux. On the other hand, when $F_0<\frac{2k}{L^2}$ there is a maximum value for the magnetic flux beyond which the configuration is no longer stable, and reduces to $k+\frac{F_0{\cal N}}{2}$ free quarks. As in previous sections this is the value for which $\beta=0$, which in this case is:
\begin{equation}
{\cal N}_{\rm max}=\frac{kL^2}{\pi(4k^2-F_0^2L^4)}\Bigl(2\pi F_0 L^2+\sqrt{F_0^2 L^4+4k^2(4\pi^2-1)}\Bigr)
\end{equation}
This behavior of the binding energy per string as a function of ${\cal N}$ and $F_0$ can be shown in Figure 8.

\begin{figure}[t]
\begin{center}
\includegraphics[scale=.5]{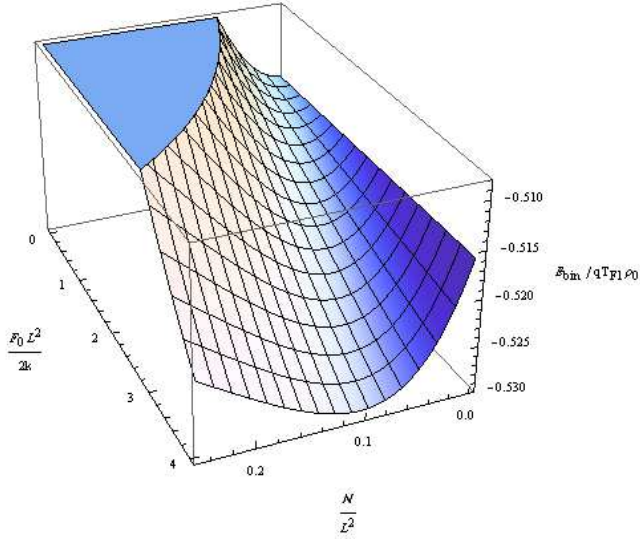}
\caption{Binding energy per string of the 't Hooft monopole, in units of $T_{F1}\, \rho_0$,  as a function of the magnetic flux and the mass.}  
\end{center}
\label{fig:EnergyD2perstringwithF0}
\end{figure}

The D4-brane with flux in the massive background has
\begin{equation}
\beta=\sqrt{1-\frac{L^4}{4\pi^4(4{\cal N}+\frac{F_0}{k}{\cal N}^2)^2}\Bigl(1+\frac{4\pi^2{\cal N}^2}{L^4}\Bigr)^2}
\end{equation}
This function has a maximum at
${\cal N}=\frac{F_0L^4}{16\pi^2 k}(1+\sqrt{1+\frac{64\pi^2 k^2}{F_0^2 L^4}})$. For this value the configuration is maximally stable. On the other hand $\beta=0$ is reached when ${\cal N}=\frac{L^2}{8\pi^2}$ for $F_0=\frac{2k}{L^2}$, and ${\cal N}=\frac{2kL^2}{2k-F_0L^2}(1\pm\sqrt{1-\frac{2k-F_0L^2}{8\pi^2 k}})$ for $F_0\ne \frac{2k}{L^2}$. For these values the configuration is no longer stable and reduces to $\frac{k{\cal N}}{2}+\frac{F_0{\cal N}^2}{8}$ free quarks. In summary the values of the magnetic flux for which the configuration can form a bound state must satisfy:
\begin{equation}
{\cal N}\ge \frac{L^2}{8\pi^2}\qquad  {\rm for}  \qquad F_0=\frac{2k}{L^2}\, ,
\end{equation}
\begin{equation}
{\cal N}\ge \frac{2kL^2}{F_0L^2-2k}\Bigl(\sqrt{1+\frac{F_0L^2-2k}{8\pi^2k}}-1\Bigr) \qquad
{\rm for} \qquad F_0>\frac{2k}{L^2}\, ,
\end{equation}
and
\begin{equation}
\frac{2kL^2}{2k-F_0L^2}\Bigl(1-\sqrt{1-\frac{2k-F_0L^2}{8\pi^2k}}\Bigr)\le {\cal N}\le
\frac{2kL^2}{2k-F_0L^2}\Bigl(1+\sqrt{1-\frac{2k-F_0L^2}{8\pi^2k}}\Bigr)
\end{equation}
for $F_0<2k/L^2$.
Note that in all cases there is a minimum value required for the magnetic flux, consistently with the fact that also in the massive case a configuration with a D4-brane and fundamental strings attached does not exist for vanishing magnetic flux.

The behavior of the binding energy per string as a function of ${\cal N}$ and $F_0$ is shown in Figure 9 (left). Figure 9 (right) exhibits the value of the magnetic flux for which the configuration is maximally stable for different values of the mass.

\begin{figure}[t]
  \begin{center}
  $\begin{array}{c@{\hspace{0.3in}}c}
    \includegraphics[scale=.5]{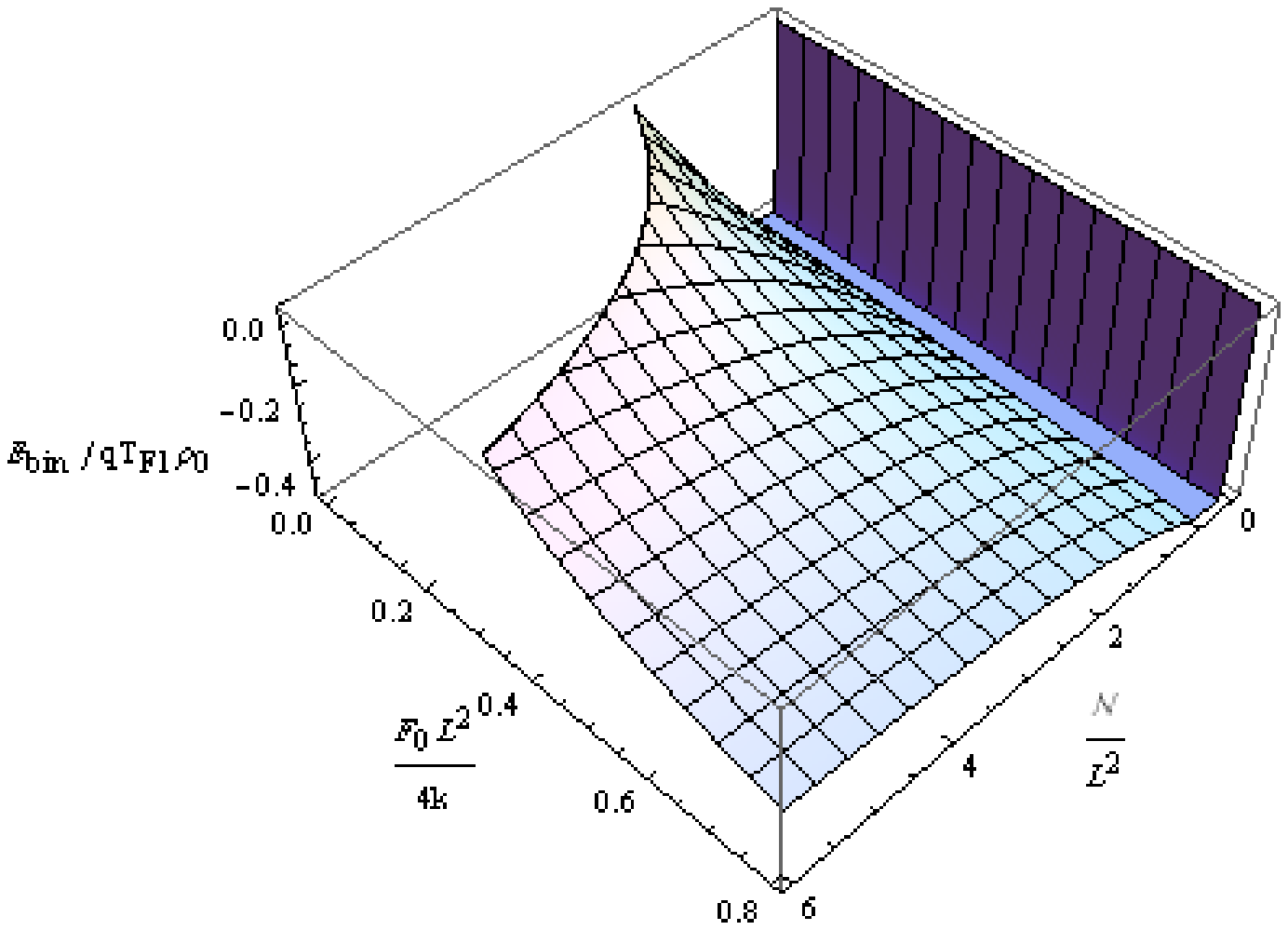}
    & \includegraphics[scale=.62]{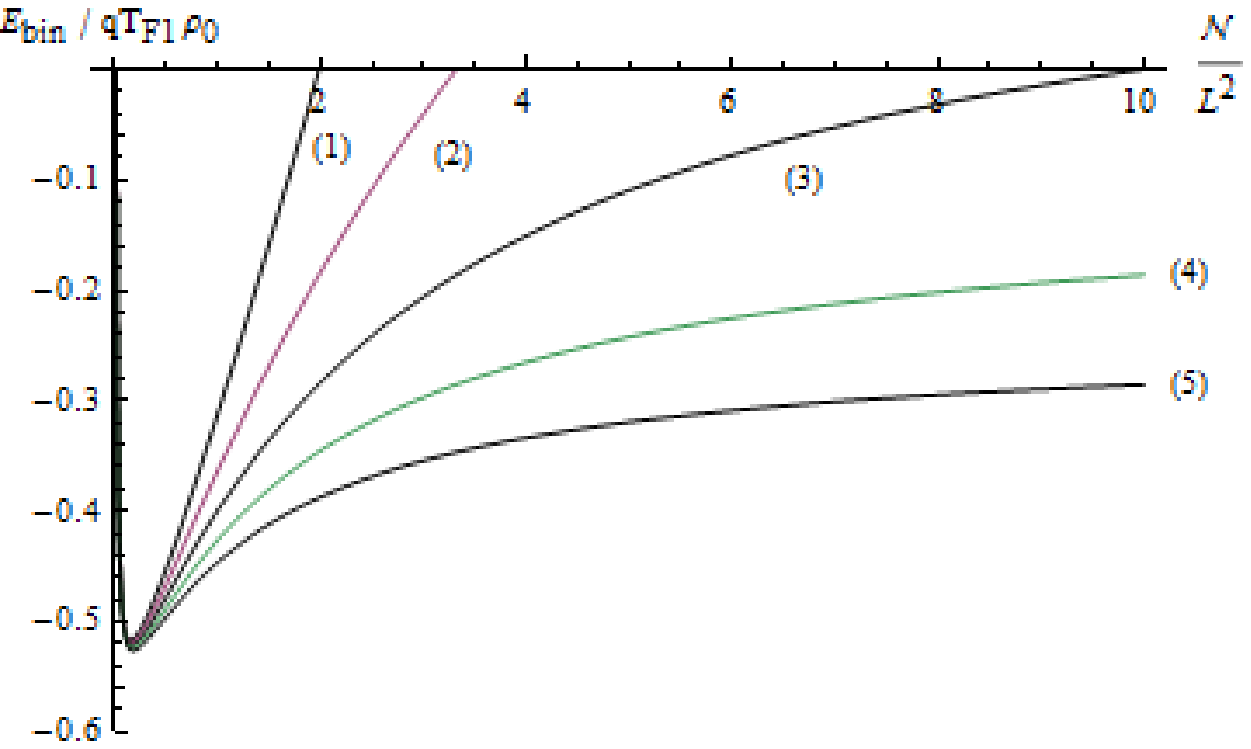}
  \end{array}$
  \end{center}
  \caption{Binding energy per string for the di-baryon, in units of $T_{F1}\, \rho_0$,  as a function
 of the magnetic flux and the mass (right: $\frac{F_0 L^2}{4k} = \left\{0,\; 0.2,\; 0.4,\; 0.6,\; 0.8\right\}$ for (1)-(5) respectively).}
  \label{fig:EnergyD4withF0}
\end{figure}

Finally, the D6-brane with flux has
\begin{equation}
\beta=\sqrt{1-\frac{1}{36\pi^2\Bigl(1+\frac{4\pi^2{\cal N}^2}{L^4}(1+\frac{F_0{\cal N}}{6k})\Bigr)^2}
\Bigl(1+\frac{4\pi^2 {\cal N}^2}{L^4}\Bigr)^3}
\end{equation}
This function reaches its maximum value when ${\cal N}=0$ for arbitrary mass. On the other hand, $\beta=0$ is reached for finite ${\cal N}$ when $F_0<\frac{2k}{L^2}$. Beyond this value of ${\cal N}$ the configuration is no longer stable and reduces to $N+\frac{k {\cal N}^2}{8}+\frac{F_0{\cal N}^3}{48}$ free quarks plus a wrapped D6-brane.

The behavior of the binding energy per string as a function of ${\cal N}$ and $F_0$ is shown in Figure 10 (left). Figure 10 (right) exhibits more clearly the behavior of the binding energy with the magnetic flux for various values of $F_0$.

\begin{figure}[t]
  \begin{center}
  $\begin{array}{c@{\hspace{0.2in}}c}
    \includegraphics[scale=.5]{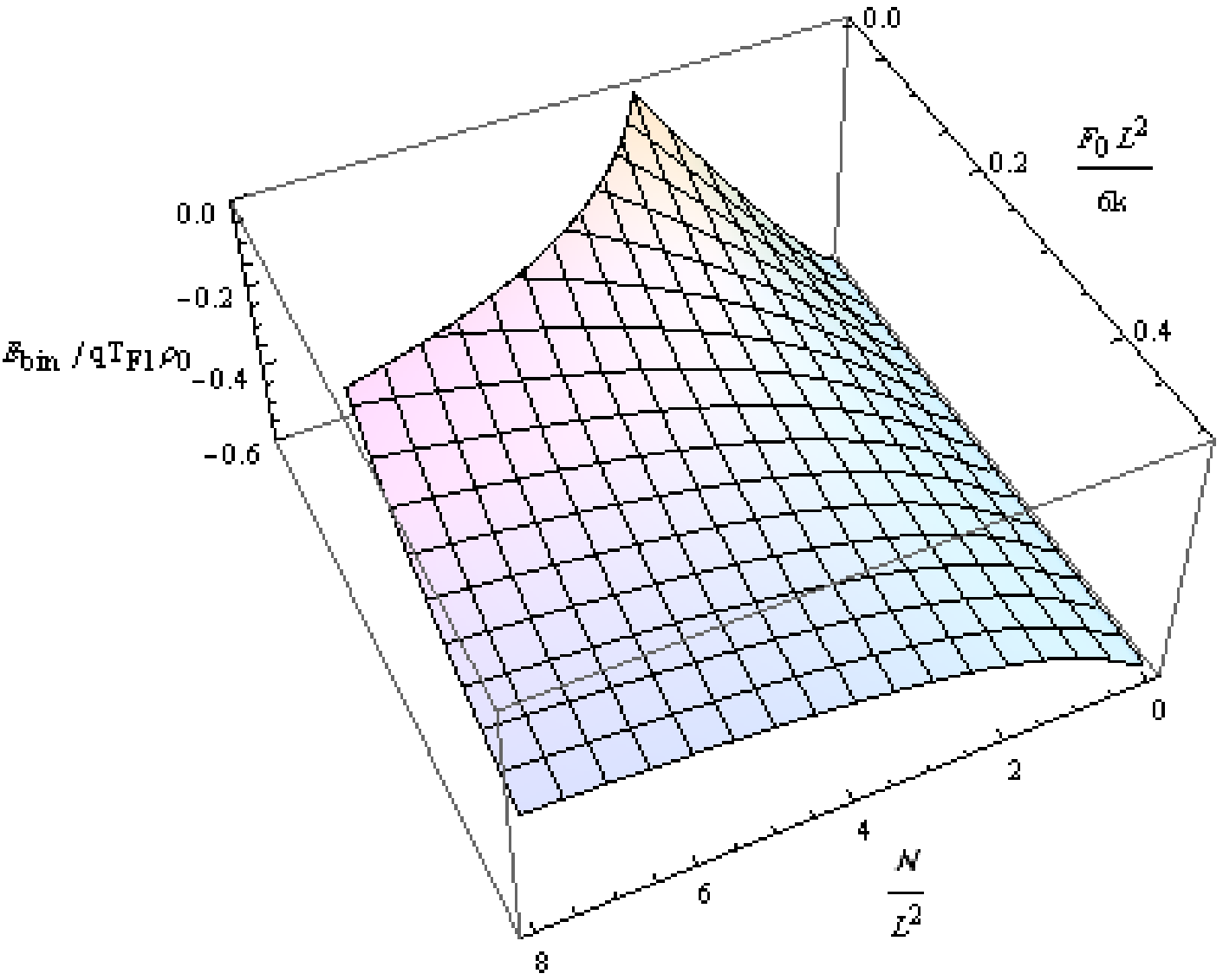}
    & \includegraphics[scale=.62]{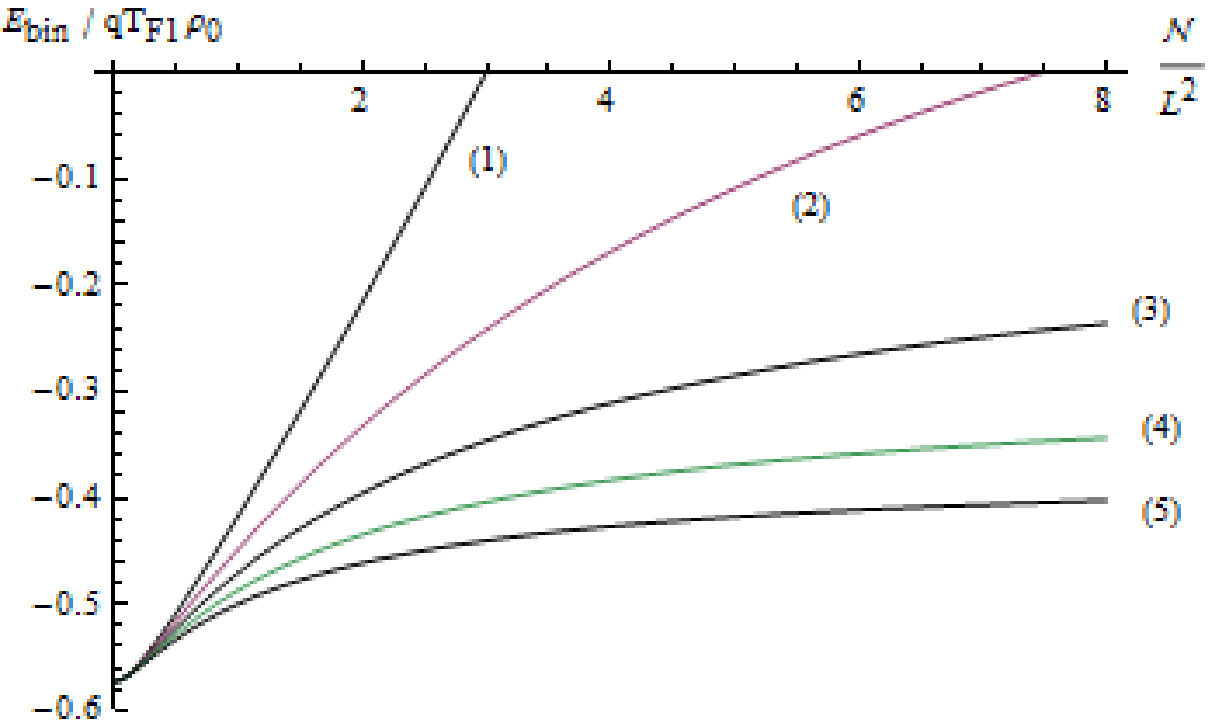}
  \end{array}$
  \end{center}
  \caption{Binding energy per string for the baryon vertex, in units of $T_{F1}\, \rho_0$,  as a function
 of the magnetic flux and the mass (right: $\frac{F_0 L^2}{6k} = \left\{0,\; 0.2,\; 0.4,\; 0.6,\; 0.8\right\}$ for (1)-(5) respectively).}
  \label{fig:EnergyD6withF0}
\end{figure}

\section{Conclusions}

In this paper we have analyzed various configurations of particle-like branes in ABJM, focusing on the study of their dynamics. This study has revealed that new and more general monopole, di-baryon and baryon vertex configurations can be constructed if the particle-like branes carry lower dimensional brane charges.

We have seen that a new di-baryon configuration with external quarks can be constructed out of the D4-brane wrapped on the $\mathbb{P}^2\subset\mathbb{P}^3$. In the presence of a non-trivial magnetic flux $F={\cal N}J$, with $J$ the K\"ahler form of the $\mathbb{P}^3$, this brane develops a tadpole that has to be cancelled with $k\,{\cal N}/2$ fundamental strings. The study of the dynamics of the D4-brane plus the 
$k\,{\cal N}/2$ F-strings reveals that the configuration is stable for $1-\sqrt{1-\frac{1}{4\pi^2}}\leq \frac{{\cal N}}{L^2}\leq 1+\sqrt{1-\frac{1}{4\pi^2}}$. Dynamically, the upper bound  arises because if the energy of the D4-brane with flux is too large the F-strings cannot form a bound state with it. For this value the strings become radial, and the configuration reduces to free $k\,{\cal N}_{{\rm max}}/2$ quarks plus the charged D4-brane. We have found as well a minimum value for the magnetic flux, that has to do with the fact that if the magnetic flux is too small the number of F-strings ending on the D4-brane is not enough to form a bound state. The existence of this lower bound was expected in this case given that the 
whole configuration of D4-brane with fundamental strings attached can only exist in the presence of flux.  
When this value is reached the configuration reduces to free $k\,{\cal N}_{{\rm min}}/2$ quarks plus a D4-brane. It is perhaps significant that the value of the magnetic flux for which the configuration is maximally stable is that for which the (off-shell) energy of the ${\cal N}^2/8$ D0-branes dissolved in the D4-brane equals the (off-shell) energy of the D4-brane. This seems to point at some kind of degeneracy for the ground state. It would be interesting to find an explanation for this phenomenon.

The D2 and D6-brane (monopole and baryon vertex) configurations exist already for vanishing magnetic flux. Consistently, no minimum value is found in the study of their dynamics. 
In these cases the effect of the magnetic flux is to allow the construction of more general monopole and baryon vertex configurations. The simplest case is the D2-brane, for which the charge of the tadpole is not modified by the magnetic flux and the number of attached F-strings is still $k$. We have seen that the configuration formed by the bound state D2-D0 plus the $k$ F-strings is stable for  ${\cal N}/L^2 \leq \sqrt{1-\frac{1}{4\pi^2}}$, reducing to $k$ free quarks plus a D2-brane with $\frac{L^2}{2}\,\sqrt{1-\frac{1}{4\pi^2}}$ D0-brane charge
when  the upper bound is reached. The D6-brane in turn is the analogue of the baryon vertex in $AdS_5\times S^5$ \cite{Witten}. The generalization of the later to include a non-vanishing magnetic flux was studied in \cite{JLR}. In that reference it was found that the magnetic flux had to be bounded from above in order to find a stable configuration, like for the D2 and D6 branes considered in this paper.
For the D6-brane the number of F-strings depends as well on the magnetic flux, but this fact does not modify substantially its dynamics. 

As we have mentioned, all the configurations that we have considered reduce to free quarks when the magnetic flux reaches the highest possible value (also the lowest for the D4-brane). For this value the brane can be located at an arbitrary position in $AdS$, with the free radial strings stretching from there to $\infty$. This is different from the free quark configuration of  \cite{BISY}, where the D5-brane is located at $\rho_0=0$, where it has zero-energy, and the radial strings stretch from 0 to $\infty$. For the maximum (and minimum, if applicable) value of the magnetic flux the D-brane is located at an arbitrary $\rho_0$, where it has some energy which is compensated by the lower energy of the strings stretching between $\rho_0$ and $\infty$. In the presence of magnetic flux the location of the D$p$-brane has therefore become a moduli of the system.  

We have already stressed that in the probe brane approximation used in this paper all supersymmetries are broken. However, in analogy with the baryon vertex construction in $AdS_5\times S^5$ we expect that, at least when the charged branes are supersymmetric, some supersymmetries will be preserved when the strings join the brane at the same point. In this case we would have to consider the full DBI problem and look for spiky solutions \cite{CGS,GRST}.
The description of the baryonic brane in $AdS_5 \times S^5$ in terms of a single D5-brane developing a spike was done in \cite{Imamura2}. This configuration is BPS, and this is reflected in the fact that its binding energy is zero. An attempt to find similar spiky solutions in $AdS_4\times \mathbb{P}^3$ has been made recently in \cite{KL}, with rather negative results even for the D6-brane with zero magnetic flux, which should be analogous to \cite{Imamura2}. We hope that some spiky configurations can still be found in this background by relaxing some of the ansatze taken in \cite{KL}. We will report on these issues in a future publication.

It is significant that for all the configurations that we have discussed the binding energy is non-analytic in the 't Hooft coupling, with this non-analyticity being of the precise form of a square-root branch cut, like in $AdS_5\times S^5$. This hints at some kind of universal behavior based on the conformal symmetry of the gauge theory. 

An important question that remains open is what are the field theory realizations of the D$p$-branes with charge that we have considered. Since we do not expect that the D2 and D6 brane configurations are supersymmetric it is hard to have an intuition about the interpretation of the new charges in the field theory side. It is interesting to note that the number of extra fundamental strings required to cancel the worldvolume tadpole is that required to cancel the tadpole on the dissolved D2 branes. This might suggest that the dual operators are doped versions of the original ones with an operator representing the D2 branes. It is hard to be more precise, in particular due to the expected lack of SUSY. However, for the D4-brane with D0-charge one can expect that a supersymmetric spiky solution exists, in which case it makes sense to try to interpret the bounds on the magnetic flux in the gauge theory dual.  In field theory language the bound (\ref{bound}) would read:
\begin{equation}
\label{boundCFT}
N+\frac{{\cal N}^2}{8}\, k \le 2\pi\, n_f \sqrt{2\lambda}\, ,
\end{equation}
 where $n_f$ is the number of external quarks, which is a function of the magnetic flux: $n_f=k\,{\cal N}/2$, and $\lambda$ is the 't Hooft coupling. Therefore, at strong 't Hooft coupling we expect a bound on the baryon charge of (generalized) di-baryon configurations with $n_f$ external quarks. This should be related in some way to the stringy exclusion principle of \cite{MS}, although we have not been able to find a direct interpretation. Note that for all branes the bound on the magnetic flux exhibits the same non-analytic behavior with $\lambda$ as the binding energy, which seems to indicate that the bounds should have its origin in the conformal symmetry of the gauge theory.
 
 Finally, the role of the $B$ field, and its potential relation to the Abelian part of the gauge symmetry, remains to be understood. We postpone these investigations for further work.

\subsection*{Acknowledgements}

We would like to thank Joan Sim\'on for useful discussions and the anonymous referee for pointing out some important numerical errors in a previous version of this paper. D.R-G. would like to thank N. Benishti and J. Sparks for collaborations on related topics and many enlightening conversations about physics and mathematics. N.G. wants to thank the CERN TH-division, where part of his work was done, for the kind hospitality. The work of N.G. was supported by a FPU-MICINN Fellowship from the Spanish Ministry of Education and the European Social Fund. This work has been partially suported by the research grants MICINN-09-FPA2009-07122, MEC-DGI-CSD2007-00042 and FICYT-IB09-069.

\end{document}